\documentclass[letterpaper, 10 pt, conference]{ieeeconf}
\IEEEoverridecommandlockouts                              
\usepackage{graphicx} 
\usepackage{subfigure}
\usepackage{textcomp,gensymb,amsmath}
\usepackage{adjustbox}
\usepackage{cite}

\title{\LARGE \bf
Analysis of Unprotected Intersection Left-Turn Conflicts based on Naturalistic Driving Data
}

\author{Xinpeng Wang$^{1}$, Ding Zhao$^{2}$, Huei Peng$^{3}$ and David J. LeBlanc$^{2}$
\thanks{* This work is funded by the Mobility Transformation Center Denso Tailor Project at the University of Michigan with grant No. N020210.}
\thanks{$^{1}$X. Wang is with the Department of Automation, Tsinghua University, Beijing, China, 100084, and now he is a visiting scholar at the University of Michigan, Ann Arbor, MI 48109, U.S. 
        }%
\thanks{$^{2}$D. Zhao (corresponding author: {\tt\small zhaoding@umich.edu}) and D. J. LeBlanc are with the University of Michigan Transportation Research Institute, Ann Arbor, MI 48109, U.S.}%
\thanks{$^{3}$H. Peng is with Department of Mechanical Engineering at the University of Michigan Transportation Research Institute, Ann Arbor, MI 48109, U.S.}%
}

\begin{document}

\maketitle
\thispagestyle{empty}
\pagestyle{empty}

\begin{abstract}
Analyzing and reconstructing driving scenarios is crucial for testing and evaluating highly automated vehicles (HAVs). This research analyzed left-turn / straight-driving conflicts at unprotected intersections by extracting actual vehicle motion data from a naturalistic driving database collected by the University of Michigan. Nearly 7,000 left turn across path - opposite direction (LTAP/OD) events involving heavy trucks and light vehicles were extracted and used to build a stochastic model of such LTAP/OD scenario, which is among the top priority light-vehicle pre-crash scenarios identified by National Highway Traffic Safety Administration (NHTSA). Statistical analysis showed that vehicle type is a significant factor, whereas the change of season seems to have limited influence on the statistical nature of the conflict. The results can be used to build testing environments for HAVs to simulate the LTAP/OD crash cases in a stochastic manner.
%
%
%
\end{abstract}
\section{INTRODUCTION}
\begin{figure}[t]
          \includegraphics[width=0.8\linewidth]{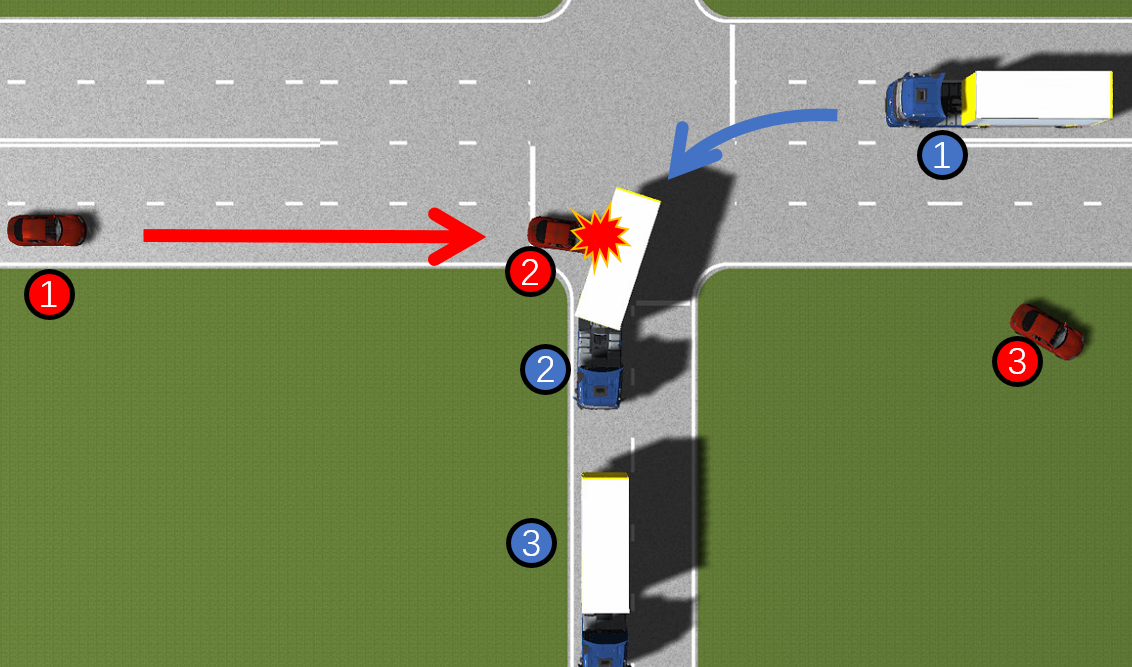}
          \centering
          \caption{A brief description of the Tesla accident}
          \label{fig:LTAPOD}
    \end{figure}

Before highly automated vehicles (HAVs) can be released to the general public, a well-defined process for testing and evaluating them must be established. The Google self-driving car project experienced 
its first shared-responsibility crash in February 2016 \cite{GoogleAutoLLC2016}. Moreover, Tesla Autopilot failed to detect a semi-truck in its first fatal crash happened in May 2016 and was criticized for using the consumers as beta testers \cite{TheinsideofaTeslavehicleatashowroominRedHook}. Fig. \ref{fig:LTAPOD} briefly demonstrates how this crash happened, with the red sedan representing the Tesla. The National Highway Traffic Safety Administration (NHTSA) is now considering the possibility of putting a pre-market approval process into place \cite{U.S.DepartmentofTransportation2016}
, in addition to a rigorous self-certification process still anticipated from the vehicle manufacturers. 

A key factor in HAV testing is the test scenarios and behaviors of other road users, particularly those of other vehicles. The test conditions need to be not only realistic but also feasible for repeated safety tests. Test scenario models can be divided into two types. The first type has fixed scenarios, such as the tests of lane support systems (LSS) \cite{NCAP2015} and autonomous emergency braking (AEB) \cite{NCAP2015a} launched by The European New Car Assessment Programme (EURO NCAP). A major advantage of this type is that it is repeatable. However, it is hard to use this type of models to represent the highly complex and variable nature of the human driving environment. Moreover, HAVs could be adjusted to pass certain fixed scenarios while their performance under broad conditions might not be well assessed. To overcome these drawbacks, we proposed a second type of models. In our previous works  \cite{Zhao2016m,Zhao2016n,Huang2016UsingScenario}, we proposed a stochastic test method and built a test environment for car-following and lane-change scenarios. In this paper, we will focus on the intersection scenario.
%
%

The intersection has been one of the most challenging scenarios for HAVs, due to the variety of road users, complexity of traffic flow and the unpredictability of vehicles and pedestrians. According to \cite{Chan2006b}, crashes at intersections took up a major portion, about 44 $\%$, of all the traffic crashes in the US. Among all kinds of scenarios with potential risks at an intersection, unprotected left turn across path - opposite direction (LTAP/OD) is a typical one. This scenario is ranked second among 10 priority V2V light-vehicle pre-crash scenarios \cite{WassimG.NajmSamuelToma2013a}.
In an LTAP/OD scenario, two vehicles are considered: the turning vehicle (TV) and the straight-driving vehicle (SdV).
    
%
%
Although a lot of research, such as \cite{Chan2006a,Nobukawa2009b}, has been conducted on traffic conflict analysis of LTAP/OD scenario, the factor of vehicle type has not been widely investigated. Now that the crash of the Tesla with Autopilot system has been attributed to its failure to detect the truck turning ahead \cite{2016PreliminaryHwy16Fh018}, it is crucial that more attention should be paid to scenarios involving heavy trucks. Moreover, there has been insufficient research on the influence of season change on driving behaviors at intersections. As extreme weather such as storm and fog has a strong impact on the driving behaviors of human drivers, we propose that they are possibly influential to HAVs as well.

This research focused on two major tasks: first, it built a stochastic model of traffic conflicts in the LTAP/OD scenario based on naturalistic driving data. Events involving both light vehicles (LVs) and heavy trucks (HTs) as the SdV were extracted from the database, reconstructed into realistic trajectories of TVs and SdVs, and finally described with several key variables. Secondly, the influence of vehicle type of the SdV and the season factor to the driving behavior of the TV was analyzed by comparing the distribution of these key variables between LVs and HTs as well as between summer and winter. 
\section{Data Source}
The data source for this research is the Integrated Vehicle-Based Safety Systems (IVBSS) database \cite{Leblanc2011a}, which was collected from 2009 to 2010 and maintained by the University of Michigan Transportation Research Institute (UMTRI). The database consists of two parts: LV platform and HT platform. It comes from a naturalistic field operational test (N-FOT) which is to assess the potential safety benefits and driver acceptance associated with a prototype integrated crash warning system\cite{Sayer2010,JamesR.SayerScottE.BogardDillonFunkhouserDavidJ.LeBlancShanBaoAdamD.BlankespoorMaryLynnBuonarosa2010}. As the system incorporates forward crash warning (FCW), lateral drift warning (LDW), lane-change/merge warning (LCM) and curve speed warning (CSW), non of the functions are designed to deal with LTAP/OD scenario. Thus, it is assumed in this research that whether the warning system is enabled will not affect driver behavior at LTAP/OD scenario. 

For LV platform \cite{Leblanc2011a}, 16 identical prototype vehicles were driven by 108 drivers for their own personal use for over six weeks. On each test vehicle, there is one long-distance 77-GHz radar that looks forward and six 24-GHz radars that cover the adjacent lanes as well as the area behind the vehicle. In addition, there is a vision system, an automotive-grade non-differential global positioning system (GPS) and an on-board digital map. Around 700 different channels of signals have been collected.

\begin{figure}
%
%
 \centering
 \subfigure[Light vehicle]{\label{fig:LVsensor}\includegraphics[width=0.44\linewidth]{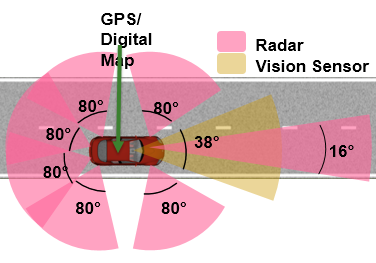}}
  \subfigure[Heavy truck]{\label{fig:HTsensor}\includegraphics[width=0.51\linewidth]{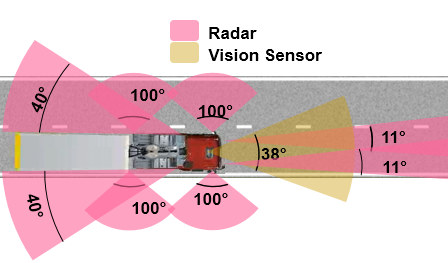}}
    \caption{Sensor configuration of IVBSS test vehicles}\label{fig:SensorConfig}
\end{figure}

For the HT platform \cite{Zhao2014d}, 18 male commercial truck drivers from Con-way Freight drove 10 equipped Class 8 tractors for 10 months. There are eight radars, three exterior cameras and several interior cameras on the test truck, recording over 500 channels of data including the driving environment, drivers’ activity, system behaviors and vehicle kinematics. Basic information about the LV and HT platforms in the IVBSS database is listed in Table \ref{tbl:IVBSS}; the configuration of on-board sensors on each platform is shown in Fig. \ref{fig:SensorConfig}.

  \begin{table}
            \centering
        \caption{INTRODUCTION TO IVBSS DATABASE}
        \label{tbl:IVBSS}
        \begin{center}
        \begin{adjustbox}{max width=\linewidth}
        \begin{tabular}{c|c c}
        \hline \hline
        Vehicle type & Light vehicle & Heavy truck\\
        \hline
        Distance & $213,309$ mi & $601,994$ mi\\
        
        Time & Apr. 2009 - Apr. 2010 &Feb. 2009 - Dec. 2009 \\
        
        Trips & 22,657 & 22,724 \\
       
        Vehicles / Drivers & 16 sedans / 108 drivers & 10 tractors / 20 drivers \\
       
        Type of front radar& Bosch LRR2 & TRW AC20 \\
        \hline \hline
        \end{tabular}
        \end{adjustbox}
        \end{center}
  \end{table}

The test area covered by IVBSS N-FOT is primarily in the Detroit area of the U.S. Most of the HT trips took place in the lower peninsula of Michigan (63 $\%$) and Ohio (33 $\%$) \cite{JamesR.SayerScottE.BogardDillonFunkhouserDavidJ.LeBlancShanBaoAdamD.BlankespoorMaryLynnBuonarosa2010}. most of the LV trips fell within a similar region \cite{Sayer2010}.

%
%
%
%
%
%
The database provides adequate information for this research. For event extraction, data from the on-board GPS sensor is used to locate the instrumented vehicle; data from the front long-range radar is used to reconstruct the trajectory of target vehicles; video recordings from vision cameras around the vehicles are used as a supplemental tool for event screening. In addition, as the IVBSS test lasted approximately one year, driving data under a variety of weather conditions throughout the year were covered, enabling us to uncover the influence of season factors. 


\section{Extraction of Left Turn Scenario }
In order to extract eligible left-turn events from the database for both the LV and the HV platform, three major tasks were performed. First, we processed data from radar for further use. Second, we searched the database for all left-turn events that meet our criteria. Finally, data points in each event were interpreted into trajectories of the TV and the SdV.

\subsection{Target Association of Truck Data}

For radar data from the HT platform, we need to associate and mark data points together that belong to the same target in order to screen out unfit targets and create a trajectory for every eligible TV. 
To cluster points of interests, we apply the following criteria to processing HT data:

\begin{itemize}
\item Only objects (TVs) that move in the opposite direction are detected. ($v_{TV} <$-0.3 m/s).
\item Only points with small azimuth angle ($|\alpha| <$ 5.5$\degree$) are considered, as the effective detecting range of the radar is 11$\degree$.
\item When the cluster with point $i$ is expended, only data points within a small time slot ([$t(i),t(i)+0.85$ s]) are considered.
\item Only neighbor points that satisfy the following rules are grouped:
\begin{enumerate}
\item Strong correspondence between range, range rate and time difference: 
\begin{equation}
|1-\frac{\delta t_{pred}}{t(j)-t(i)}| < 0.3
\end{equation}
where 
\begin{equation*}
\delta t_{pred} = 2*\frac{r(j)-r(i)}{rr(j)+rr(i)}
\end{equation*}
Here, $r(i)$ is the range of point $i$, and $rr(i)$ is the range rate of point $i$
\item Reasonable difference in transversal. 
\begin{equation}
\frac{tr(j)-tr(i)}{t(j)-t(i)}< 20   m/s
\end{equation}
 Here, $tr(i)$ is the transversal of data point $i$, and $t(i)$ is the time.
\end{enumerate}
\end{itemize}

\begin{figure}
%
%
 	\centering
     \subfigure[Range]{\label{fig:RangeCl}\includegraphics[width=0.47\linewidth]{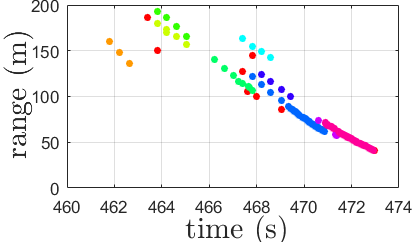}}
  \subfigure[Transversal]{\label{fig:TransversalCl}\includegraphics[width=0.47\linewidth]{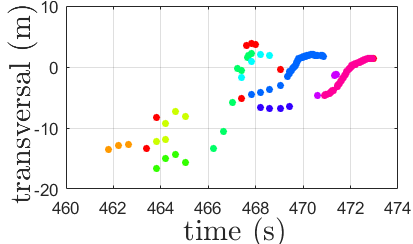}}
      \caption{An exemplary result of target association}
      \label{fig:Clustering}
\end{figure}

Fig. \ref{fig:Clustering} shows an example of data points that are associated and divided into different groups in one event. The dots with the same color show trajectories of targets, while red dots do not belong to any group and are seen as noise. In such a typical LTAP/OD scenario, a vehicle is turning in front of the instrumented truck. Fig. \ref{fig:RangeCl} shows how the range of target points change over time. As target vehicles cross the intersection when the instrumented truck is moving forward at a steady speed, the ranges to different targets are decreasing linearly. Moreover, Fig. \ref{fig:TransversalCl} shows that the transversal of multiple targets is increasing from negative to positive, indicating that they cross from left to right in the view of the instrumented truck. Once data points from each target are clustered, the HT platform can be used for further event extraction for eligible LTAP/OD scenarios. 
\subsection{Event Screening}
\begin{figure}
   \includegraphics[width=45mm]{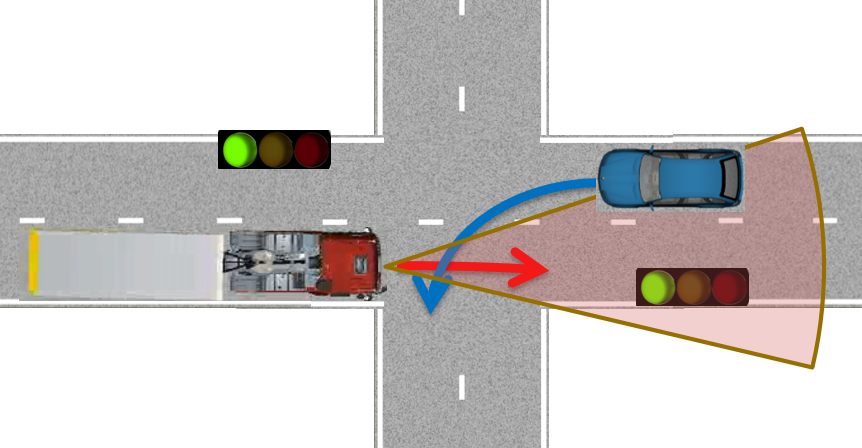}\quad
   \includegraphics[width=35mm]{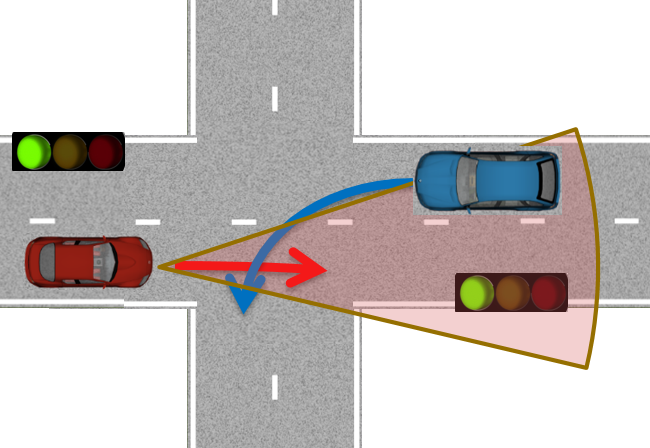}
    \caption{Configuration of the instrumented vehicle and the target vehicle for event extraction in LTAP/OD scenario}
    \label{fig:SDConflictLVHT}

 \end{figure}

An unprotected LTAP/OD scenario can be recorded by either the SdV or the TV. In this paper, we use only the scenarios recorded by SdVs. Fig. \ref{fig:SDConflictLVHT} demonstrates the configuration of the instrumented vehicle, i.e., the SdV and the target vehicle, i.e., the TV for event extraction in LTAP/OD scenarios. For both the LV and HT platforms, eligible left-turn events are queried based on the following criteria:
\begin{itemize}
\item The intersection has a stop sign or a set of signal lights. Although there will be protected LTAP/OD events retrieved with this criterion, they can be screened out by the following conditions, such as the constraint on the velocity of the TV and the SdV. 
\item The instrumented vehicle is moving straight (speed larger than 3 m/s  \&  change of heading angle smaller than 10$\degree$).
\item The target vehicle is moving towards the instrumented vehicle (the longitudinal projection of speed smaller than -0.5 m/s) and moving from left to right (due to the difference in radars, transversal goes from positive to negative for LVs, and from negative to positive for HTs).
\item Time duration of the event is adequate (more than 1.5 s).
\item The maximum of time difference between two consecutive points (defined as $\delta t$) in an event should be small enough to be seen as points of the same target ($max\{\delta t\} < 1$ s).
\end{itemize}

Event extraction follows a similar procedure for LV and HT. For the LV platform, we first select all straight-driving occurrence at intersections; we then extract those with left-turn objects from the opposite direction. These tasks are completed in the Microsoft SQL Server Management Studio (SSMS). Afterwards, the extracted events are exported to MATLAB for the last round of screening, which guarantees reasonable speed, targets, and time duration. For the HT platform, the only difference is that after retrieving all the occurrences of straight-driving at intersections, we export data from the database server directly into MATLAB for target association and the following extraction tasks.
 
\begin{figure}         \includegraphics[width=\linewidth]{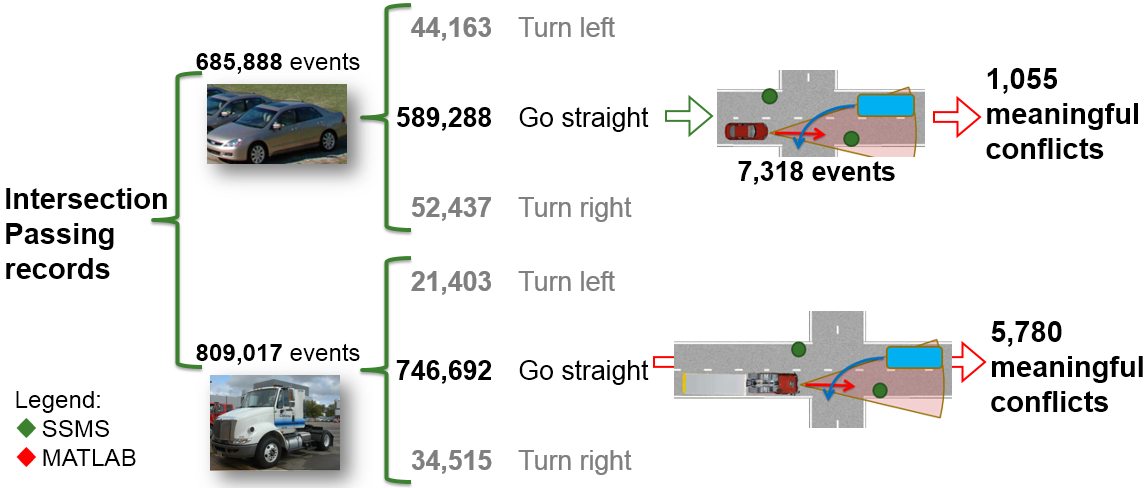}
          \caption{Procedure and interim results for event extraction}          \label{fig:eventextraction}
    \end{figure}

\begin{figure}[h!]
%
%
  \centering
       \subfigure[Light vehicle]{\label{fig:light vehicles}\includegraphics[width=0.47\linewidth]{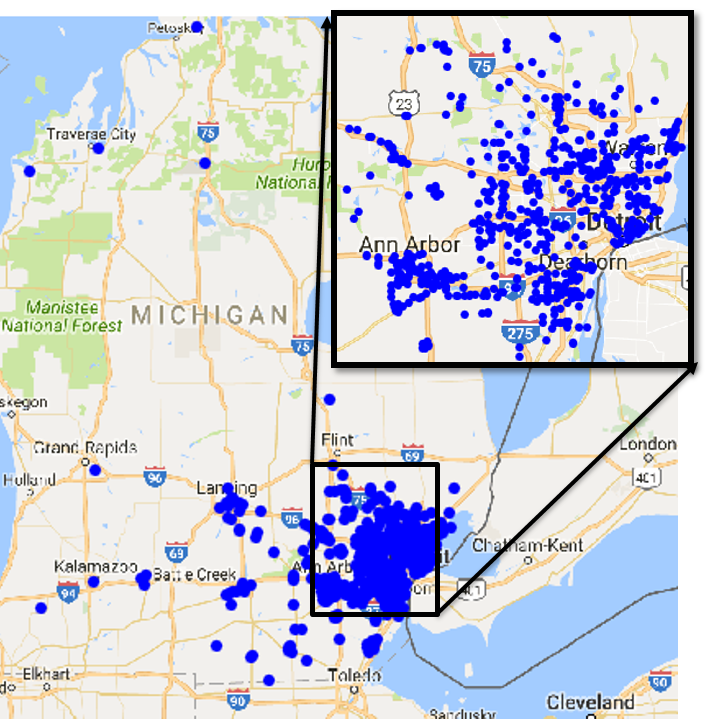}}
  \subfigure[Heavy truck]{\label{fig:heavy trucks}\includegraphics[width=0.47\linewidth]{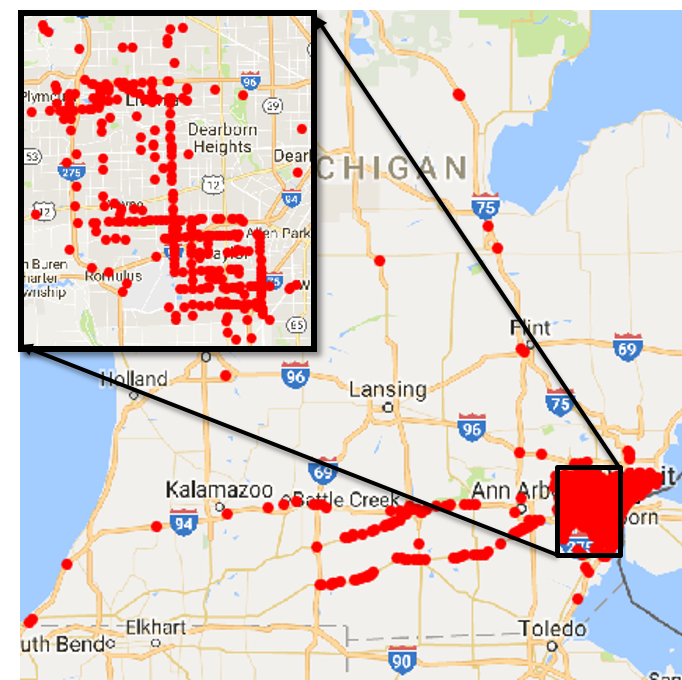}}

     \caption{The location of extracted events}
     \label{fig:RangeOfData}
 \end{figure}

The diagram in Fig. \ref{fig:eventextraction} illustrates the procedure and interim results of each phase for event extraction. 
Finally, HT has 5,780 eligible LTAP/OD events, whereas LV has 1,055. The location of these events is shown in Fig. \ref{fig:RangeOfData}.

\subsection{Trajectory Reconstruction}
\begin{figure}
  \includegraphics[width=\linewidth]{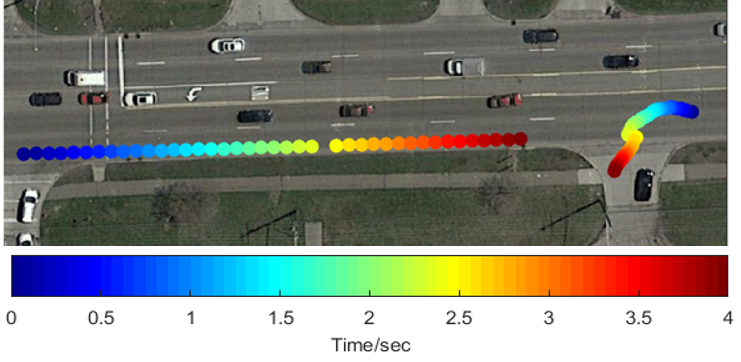}
      \caption{Example of reconstructed trajectory}
      \label{fig:SDLV_latlong}
  \end{figure}

Once all eligible events have been selected, the trajectories of both TV and SdV in each event are then reconstructed. The exact position of SdV comes from the on-board GPS sensor; the data from the front long-range radar are used to extract the relative position of TV in the coordinate of SdV. 
After synchronization on GPS and radar data, the trajectories of TV and SdV are generated. Fig. \ref{fig:SDLV_latlong} shows the reconstructed trajectories for SdV and TV in one event. Here, dots with the same color represent the position of TV and SdV at the same moment. The TV crossed intersection before the SdV in this example.  

\section{Conflict Analysis and Comparison}
\subsection{Definition and Metrics of Conflicts}

In this section, conflict is used to describe risky events in traffic. According to \cite{Tarko2012b}, conflict is defined as “an observational situation in which two or more road users approach each other in space and time to such an extent that a collision is imminent if their movements remain unchanged”. Many conflict metrics have been used for measuring the level of safety for an LTAP/OD event, including post-encroachment time (PET), leading buffer (LB) and trailing buffer (TB) used by \cite{nobukawa2011model}, and gap time (GT) used by \cite{Misener2007CaliforniaReport}. For this paper, as the goal is to construct a stochastic model, we choose only the most representative time slice in each event to model LTAP/OD conflicts. The heading angle of the SdV is taken as constant during each event, with any small deviation being ignored. Thus, a conflict point is naturally defined as the location of the TV when its transversal in the radar of SdV crosses zero, and this exact moment is regarded as the representative moment of this conflict, defined as $T_x$. Consequently, four variables at $T_x$ are chosen to model the conflict, including two modified conflict metrics: time to the conflict point ($T_{cp}$) and distance to the conflict point ($D_{cp}$):
  \begin{itemize}
  \item $D_{cp}$: Distance to the conflict point for the SdV at $T_x$
  \begin{equation}
  D_{cp}=dist(P_{SdV}(T_x),P_{TV}(T_x))
  \end{equation} Here $P_{SdV}$ and $P_{TV}$ are the positions of SdV and TV
  \item $T_{cp}$: Time to the conflict point for the SdV at $T_x$ 
  \begin{equation}
  T_{cp}=D_{cp}/v_{SdV}
  \end{equation}
  \item $v_{SdV}$: Speed of the SdV at $T_x$
  \item $v_{TV}$: Speed of the TV at $T_x$
  \end{itemize}

First, we demonstrate an example of conflict analysis on a single LTAP/OD event. Here we use the aforementioned occurrence, where the TV crossed the intersection before the SdV did.
    \begin{figure}
    \includegraphics[width=\linewidth]{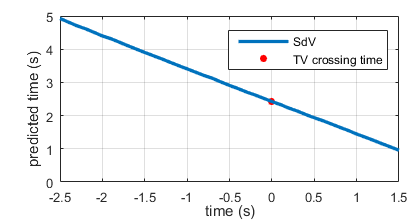}
    \caption{Time to the conflict point of the SdV in a LTAP/OD scenario}
    \label{fig:SDConflict}
    \end{figure}
Fig. \ref{fig:SDConflict} uses $T_{cp}$ to demonstrate how the SdV and the TV interacted in one real LTAP/OD event. The vertical axis indicates predicted time to the point of conflict of the SdV, whereas the horizontal axis shows the real elapsed time relative to the moment when the TV crosses the intersection, that is, $T_x$.
In this scenario, time to the conflict point decreased linearly over time, indicating the margin between the TV and the SdV was large enough for the SdV to maintain a nearly constant speed when the TV was crossing. When the TV reached the conflict point, there was a 2.3-second margin for SdV, that is, $T_{cp}$, which is demonstrated by the red dot. Here, $T_{cp}$ described the essence of this interaction between the SdV and the TV.

Then, the following modeling and analysis will ignore the detailed interaction of the TV and the SdV, paying attention only to the four aforementioned variables in each event. We use all events we retrieved in the previous section from both the HT and LV platforms as the source for modeling.
\subsection{Comparison between Light Vehicles and Heavy Trucks}
In this section, the effect of vehicle type on traffic conflict in LTAP/OD scenarios is discussed. Distributions of variables for LVs and HTs are compared. 

  \begin{figure}[b]        
   \subfigure[Distribution of $D_{cp}^{-1}$]{\label{fig:Dcp-1}\includegraphics[width=0.48\linewidth]{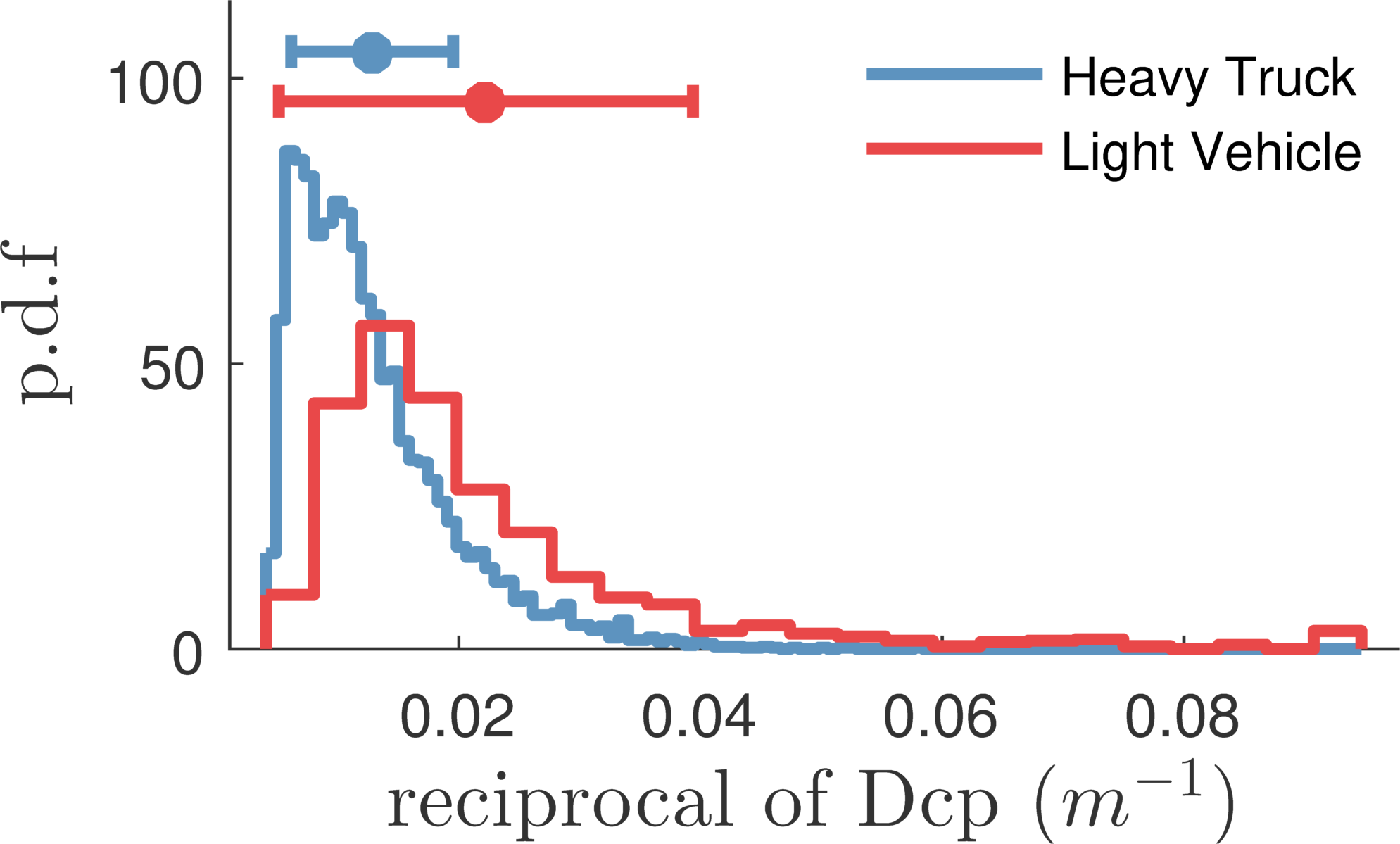}}
   \subfigure[Distribution of $T_{cp}^{-1}$]{\label{fig:Tcp-1}\includegraphics[width=0.48\linewidth]{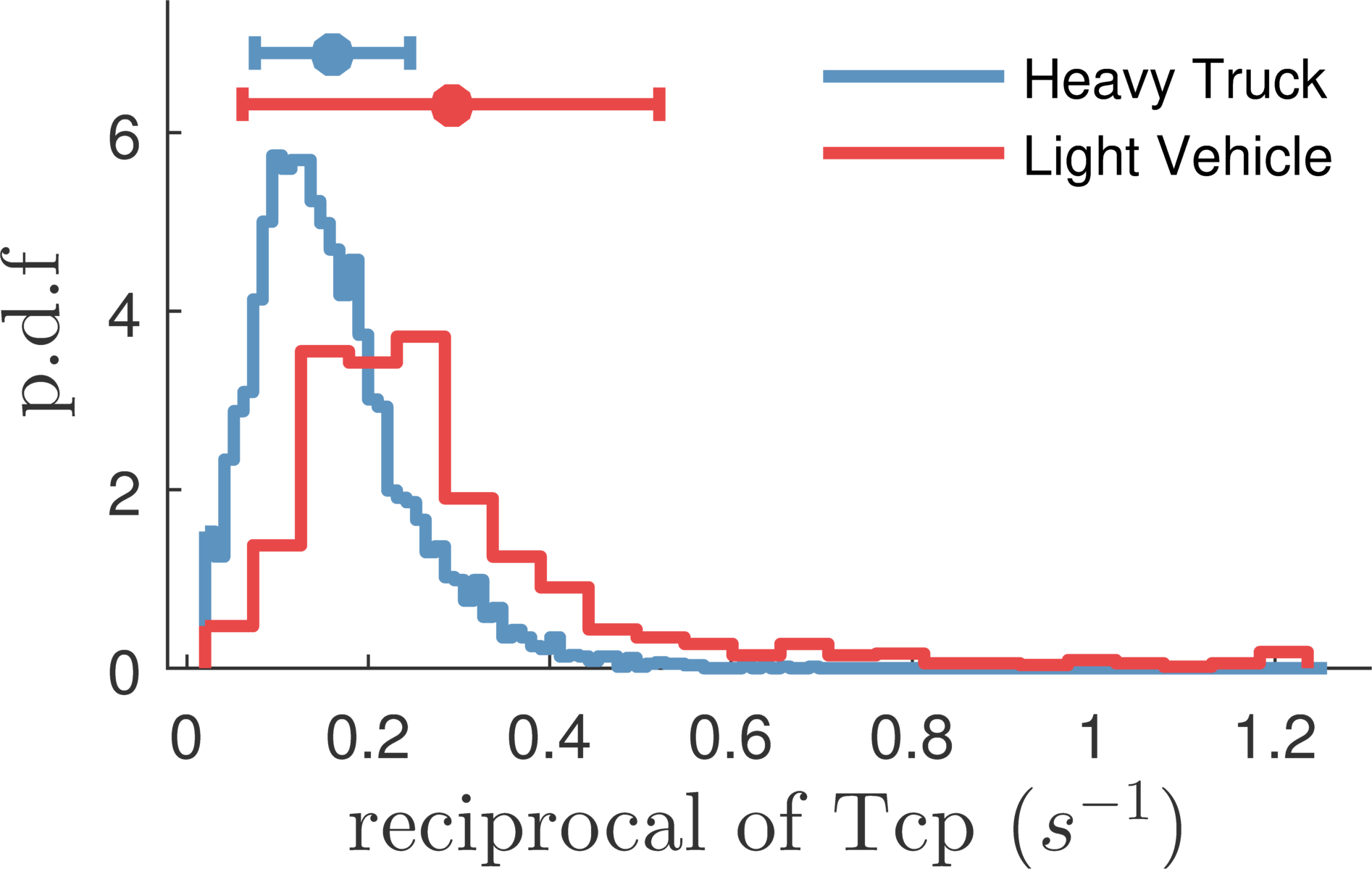}}
\caption{Comparison of $D_{cp}^{-1}$ and $D_{cp}^{-1}$ between heavy trucks and light vehicles}
     \label{fig:Tcp and Dcp}
  \end{figure}



 As events with smaller $D_{cp}$ and $T_{cp}$ are more dangerous, we generated the distributions of the reciprocal of $D_{cp}$ and $T_{cp}$  to put these risky but rare events in the tail, as shown in Fig. \ref{fig:Tcp and Dcp}. 
 The dots and bars at the top of figures show the mean value and the standard deviation of each empirical distribution. From Fig. \ref{fig:Tcp and Dcp}, we can see that when $D_{cp}^{-1}$ or $T_{cp}^{-1}$ increases, there are fewer points of data, giving rise to a shape with a long tail. Moreover, events with an HT as SdV tend to have both smaller $D_{cp}^{-1}$ and smaller $T_{cp}^{-1}$ than with an LV, indicating less severe conflicts.



\begin{figure}
  \subfigure[Speed distribution of straight-driving heavy trucks and straight-driving light vehicles]{\label{fig:SpeedSdV}\includegraphics[width=0.48\linewidth]{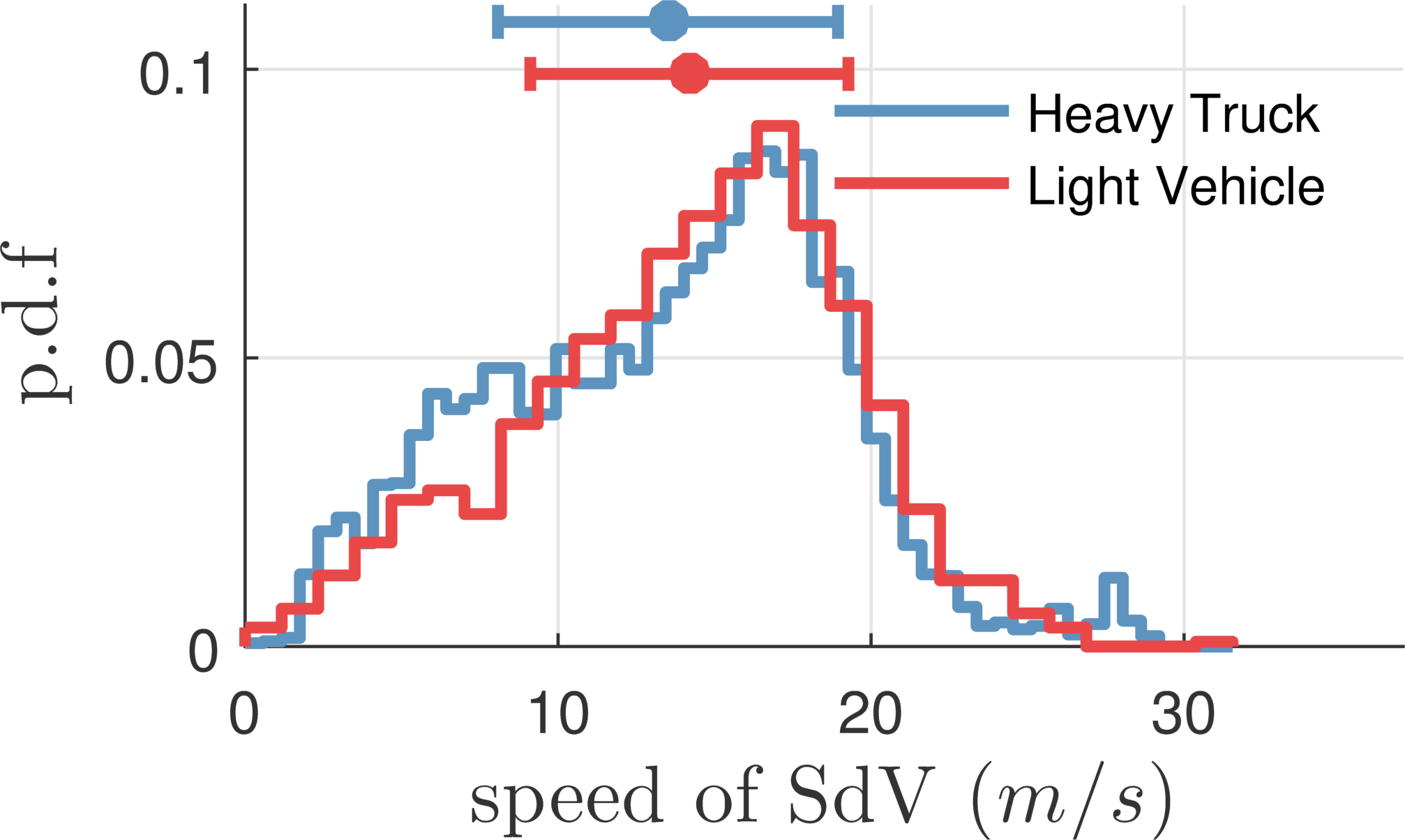}}
  \subfigure[Speed distribution of turning heavy trucks and turning light vehicles]{\label{fig:SpeedTV}\includegraphics[width=0.48\linewidth]{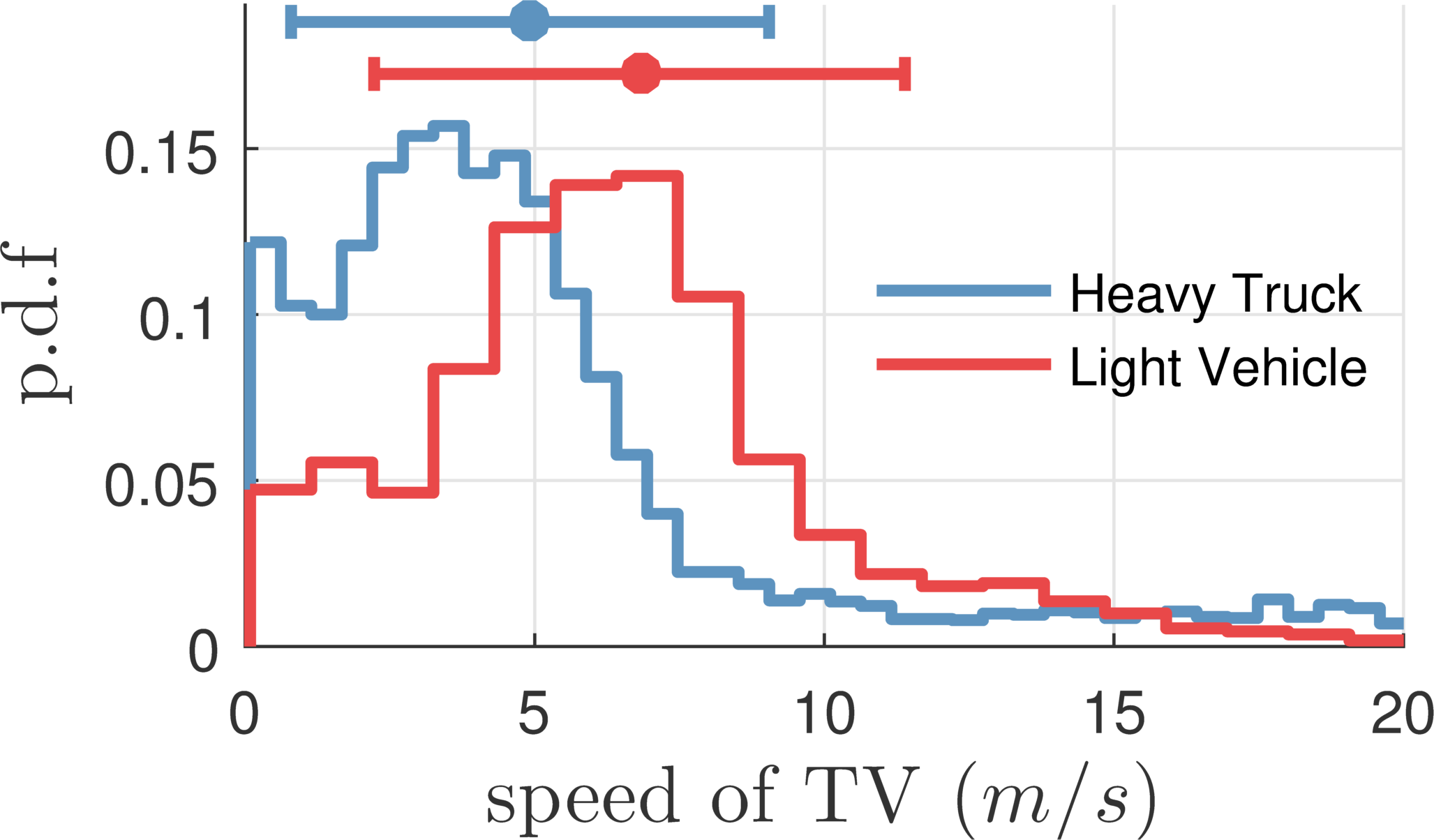}} 
      \caption{Comparison of the speed between heavy trucks and light vehicles}
      \label{fig:GroupBySpeedTcp}
 \end{figure}
Fig. \ref{fig:GroupBySpeedTcp} shows the distributions of $v_{SdV}$ and $v_{TV}$. The distribution of $v_{SdV}$ for HT and LV platforms both have a triangular shape. Most $v_{SdV}$ ranges from 12 to 20 m/s, whereas most $v_{TV}$ is less than 10 m/s at the conflict point. Though there is no obvious difference with the distribution of $v_{SdV}$ between events with LV and HT as the SdV, the $v_{TV}$ tends to be significantly higher when the SdV is an LV than is an HT. Combined with the previous results of $D_{cp}$ and $T_{cp}$, we can conclude that for left-turn conflicts where HTs are SdVs, conflict metrics have significantly higher value, and TVs tend to turn with less aggressive speed. This means that when the TV chooses the time of turning and commences turning action, it behaves more conservatively when confronted by an HT coming from the opposite direction than an LV. The difference in vehicle type does influence the driving behavior of the TV and the severity of the conflict.

 \subsection{Analysis of Season Factor}
In this section, we uncover the influence of season factor on behaviors of SdVs and TVs in LTAP/OD scenarios. During the test, 7 $\%$ of driving for HTs \cite{JamesR.SayerScottE.BogardDillonFunkhouserDavidJ.LeBlancShanBaoAdamD.BlankespoorMaryLynnBuonarosa2010} and 15 $\%$ \cite{Sayer2010} for LVs took place in freezing temperature. The months with events that took place in freezing temperatures are defined as winter, which includes December through March of the following year. This period also coincides with the time when the average snowfall in Ann Arbor is over 8 inches. On the other hand, summer is defined as being from June to August. We have retrieved 272 events in summer and 391 events in winter for LVs, whereas the numbers for HTs are 1818 and 844 respectively. $T_{cp}^{-1}$, $D_{cp}^{-1}$, $v_{SdV}$ and $v_{TV}$  are compared for  summer and winter driving. 

\begin{figure}[t]
     \centering
     
     \begin{center}
     \begin{tabular}{  c | c  }
      Light vehicle & Heavy truck\\
      \\
      \label{fig:VSdVSeasonLV}\includegraphics[width=0.48\linewidth]{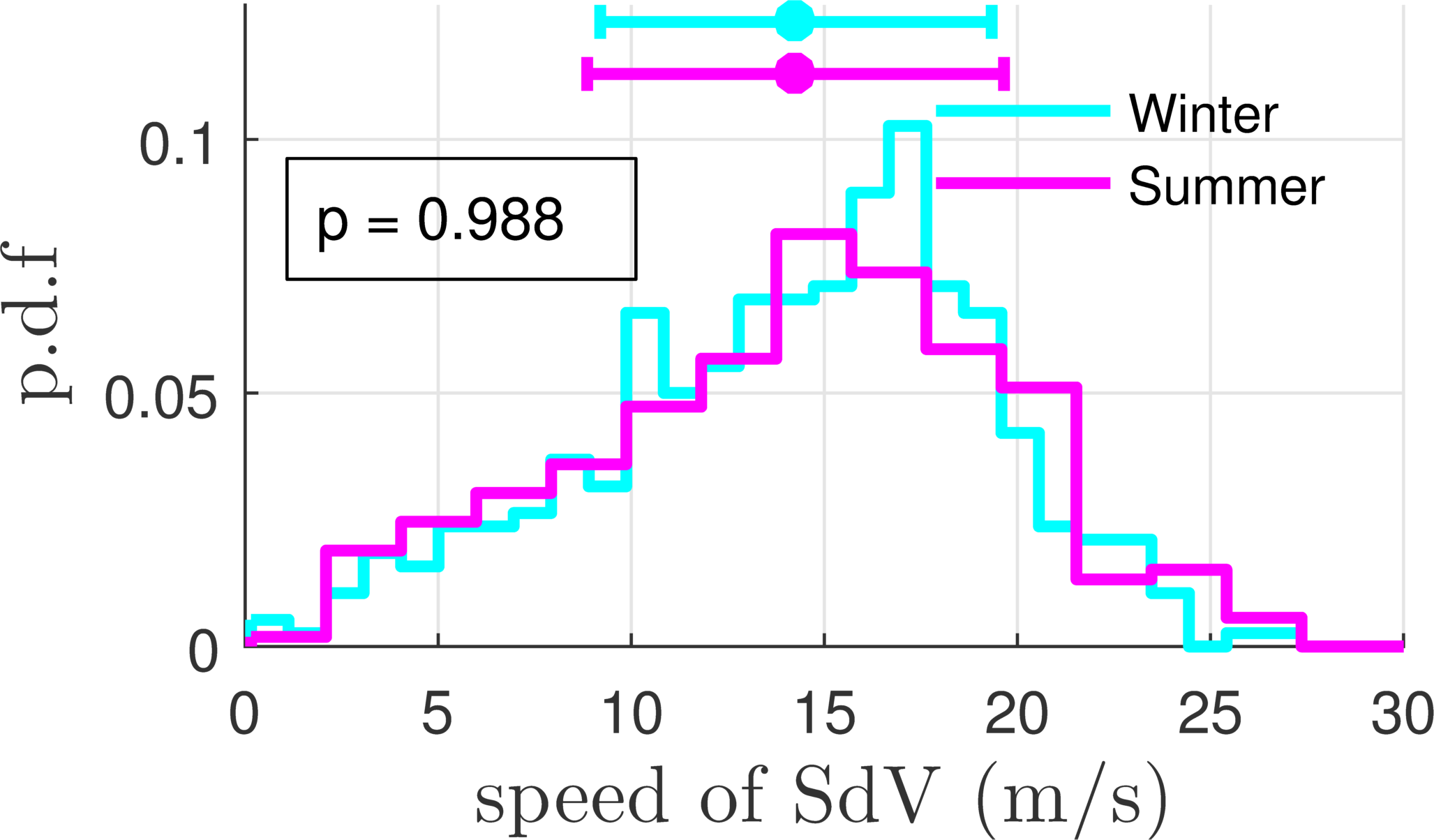}
&       \label{fig:VSdVSeasonHT}\includegraphics[width=0.48\linewidth]{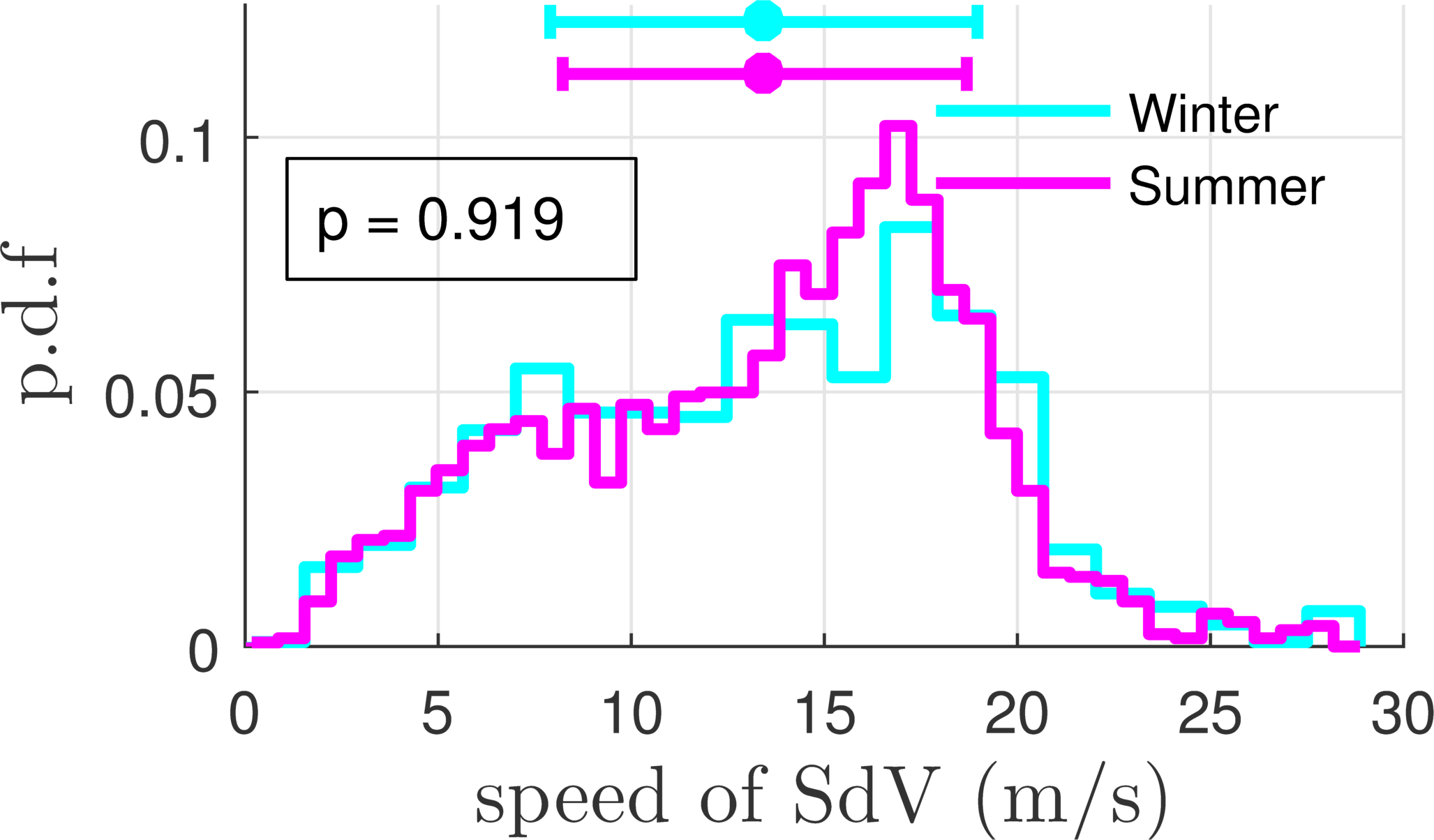}
 \\ \\
\label{fig:VTVSeasonLV}\includegraphics[width=0.48\linewidth]{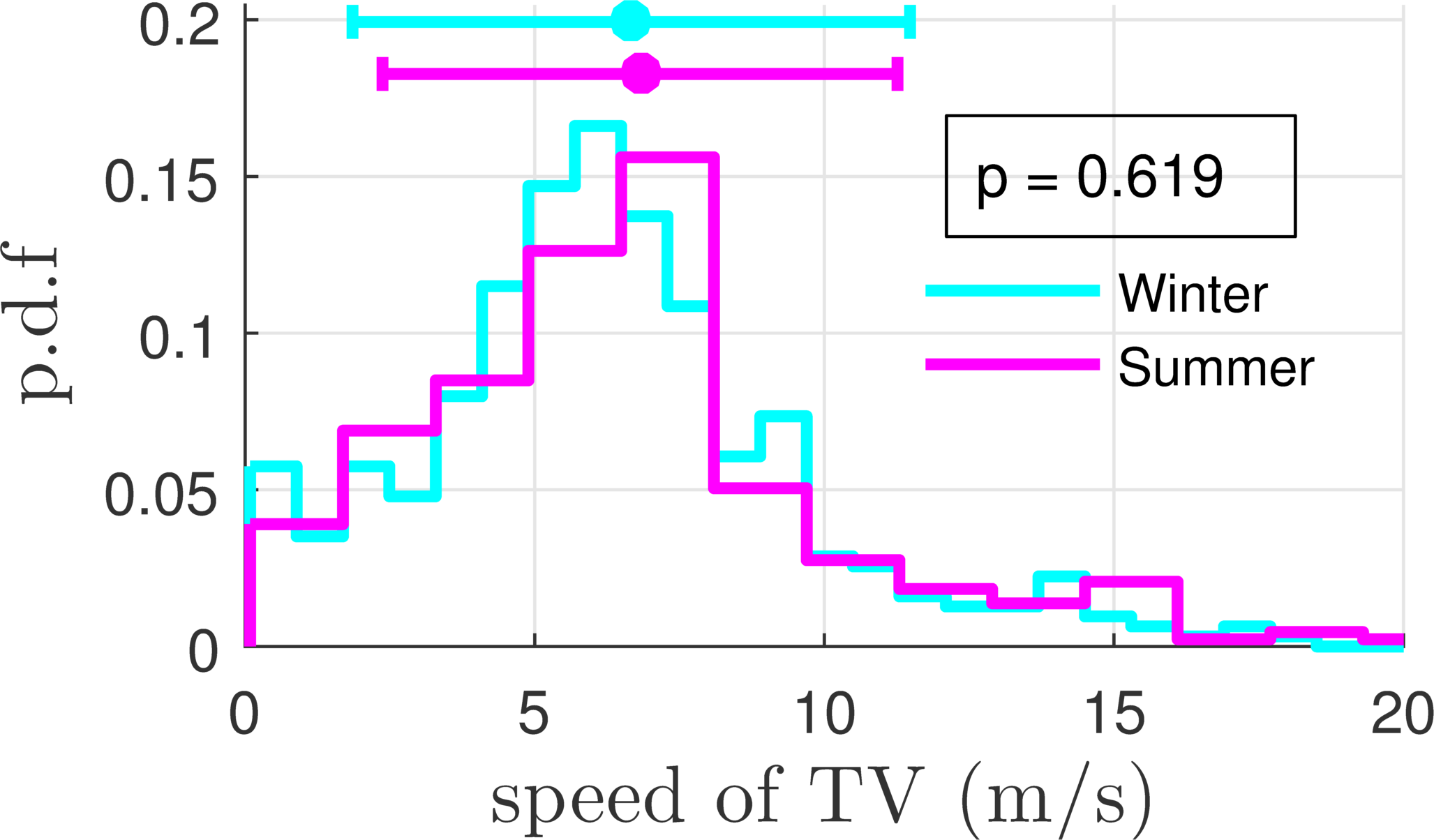}
&
\label{fig:VTVSeasonHT}\includegraphics[width=0.48\linewidth]{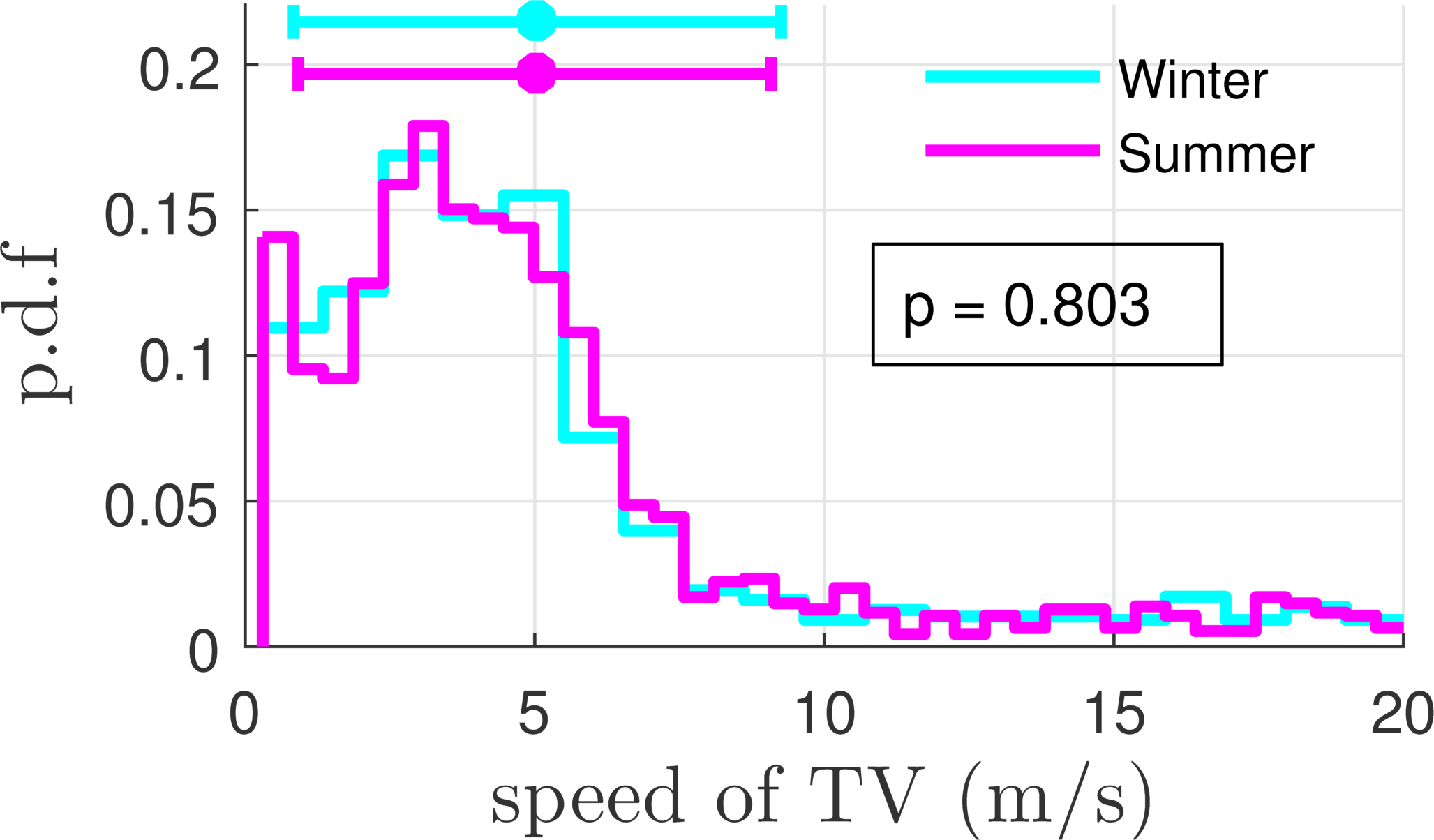}
\\
 \label{fig:DcpSeasonLV}\includegraphics[width=0.48\linewidth]{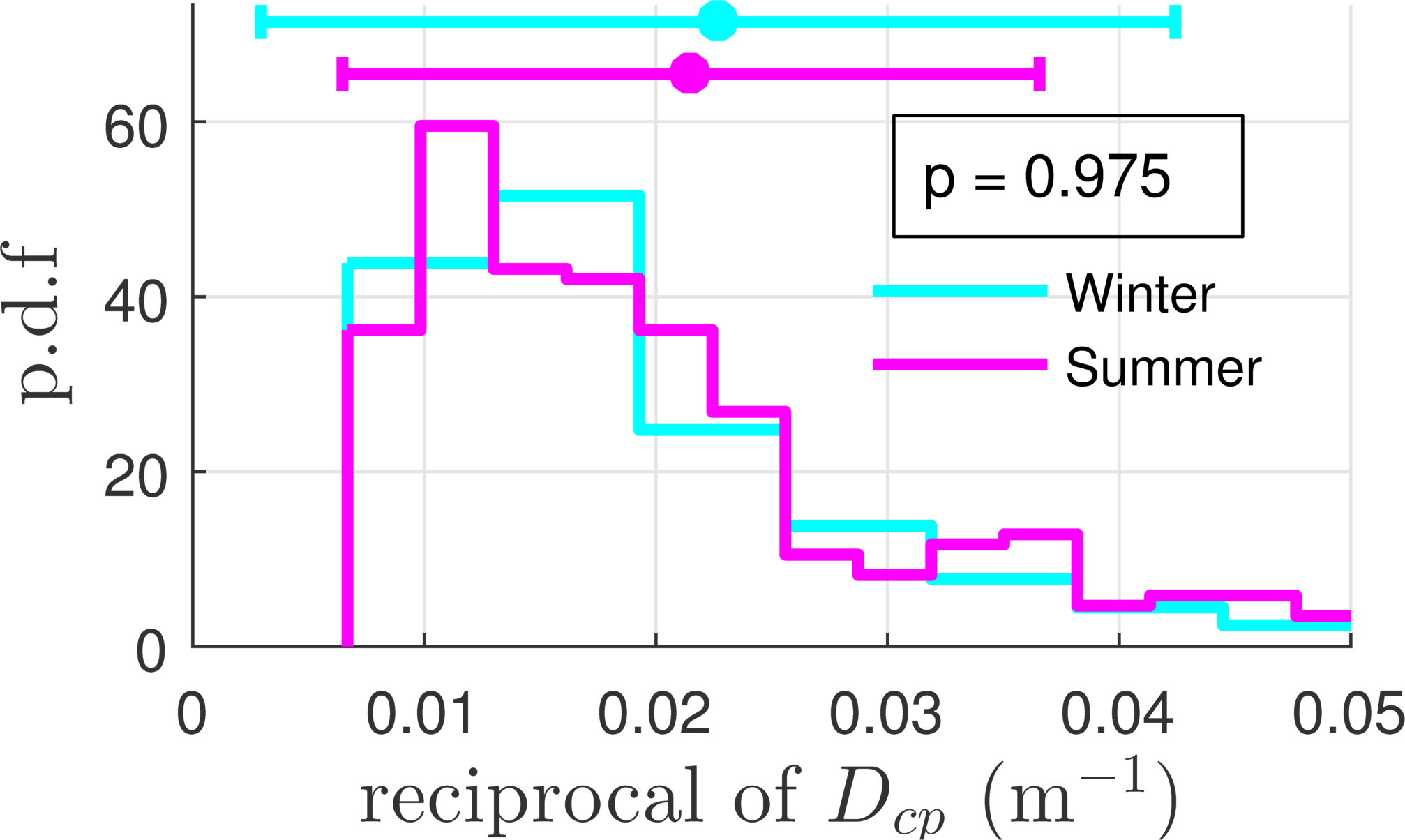}
 &
 \label{fig:DcpSeasonHT}\includegraphics[width=0.48\linewidth]{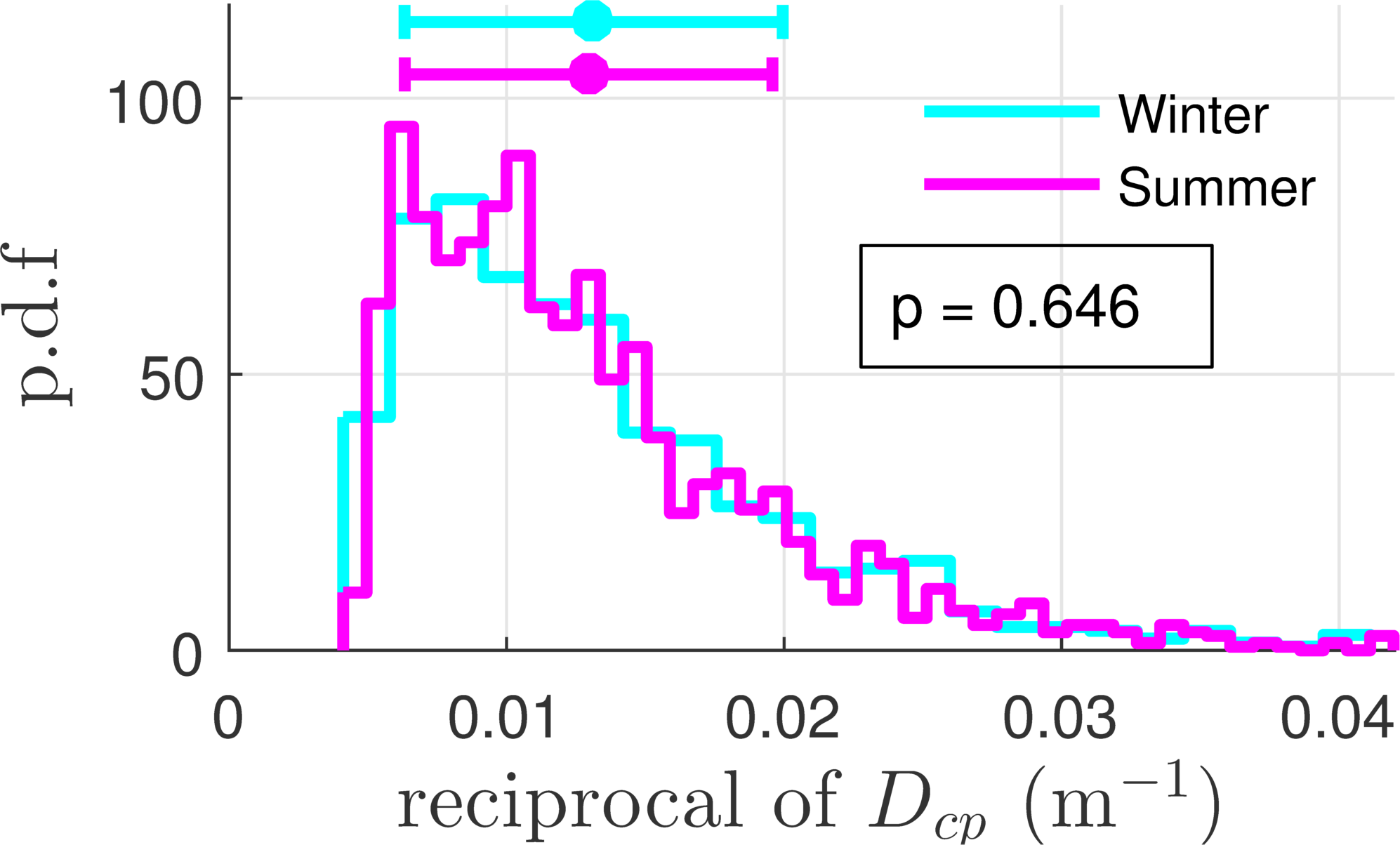}
 \\ \\
\label{fig:TcpSeasonLV}\includegraphics[width=0.48\linewidth]{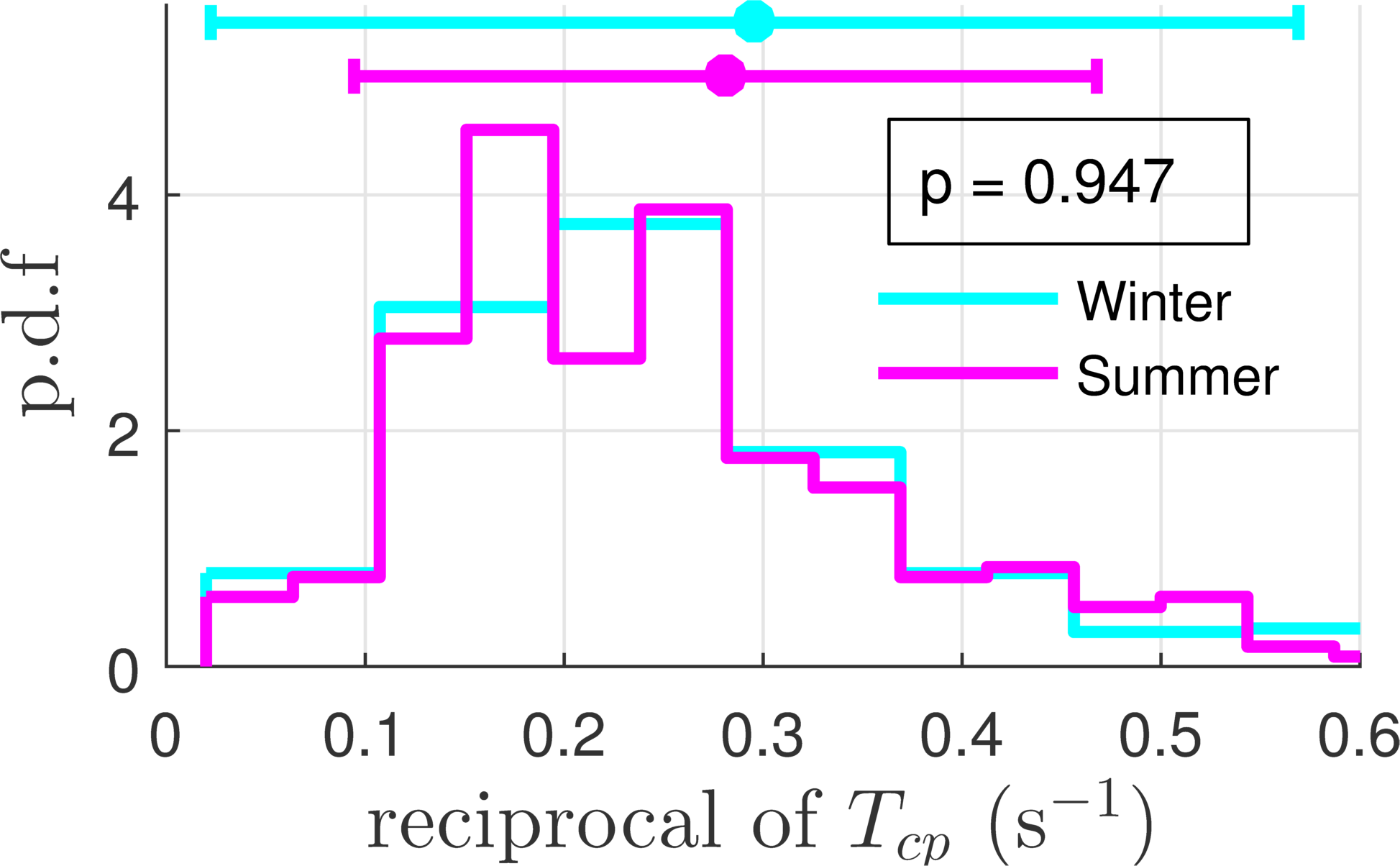}
&
\label{fig:TcpSeasonHT}\includegraphics[width=0.48\linewidth]{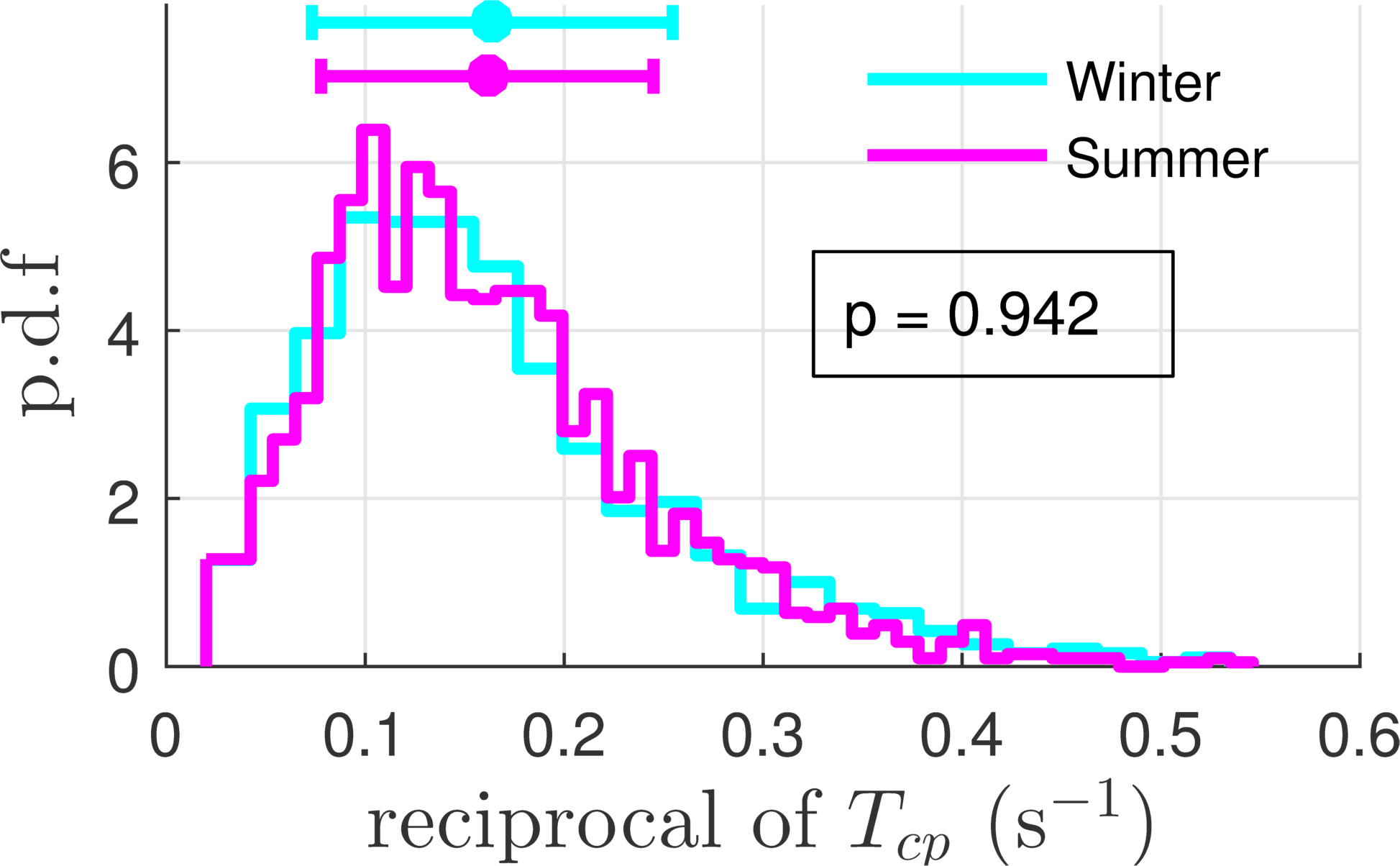}
\\ 
      \end{tabular}
      \caption{Comparison between events in summer and winter}
      \label{tbl:Season}
      \end{center}
      \end{figure}

Mann-Whitney-Wilcoxon (MWW) test \cite{Mann1947} is a non-parametric hypothesis test of the null hypothesis that two populations are the same against an alternative hypothesis. Here, we used it to determine whether the conflict metrics differ between summer and winter.

Fig.~\ref{tbl:Season} shows the result of the comparison. It can be concluded that for both LV and HT platforms, the mean values for summer and winter of all four variables that describe the conflict at LTAP/OD for both SdVs and TVs are very close. As the p-value from the MWW test is large (p-value $>$ 0.6) for all the eight distributions, we are not able to distinguish between the left-turn pattern in summer and in winter in terms of $D_{cp}^{-1}$, $T_{cp}^{-1}$, $v_{TV}$ and $v_{SdV}$. This result indicates that despite a large difference in climate, there is no significant difference between the way people drive in winter and in summer at LTAP/OD scenarios in the Great Lakes area. This conclusion has its significance for designing and testing of HAVs. 
\section{CONCLUSION}

In this research, traffic conflicts of TVs and SdVs in LTAP/OD scenarios are modeled and analyzed based on nearly 7,000 left-turn events extracted and reconstructed from the naturalistic database. The two modified conflict metrics, $T_{cp}$ and $D_{cp}$ are used to model turning behavior of the TV. This stochastic model can be further used for developing simulation tools for evaluating HAVs.

The significance of vehicle type and season are also addressed in the research. In general, when the SdV is an HT, the driver of the TV tends to turn in a more conservative fashion with a wider margin. Surprisingly, despite prevailing snow and freezing weather in the winter of Michigan, driver behavior at LTAP/OD scenarios during the N-FOT test did not differ significantly between summer and winter. These two conclusions can be useful for designing automated driving algorithms and for establishing regulations and policies for HAVs.

In the following research, we will improve the accuracy of trajectory reconstruction by conducting sensor fusion to the GPS and yaw rate sensor, and by re-synchronizing data from different channels. Moreover, we will further investigate the reasons behind the conclusion on the similarity of driver behavior between summer and winter. Possible causes could be: snow on the road was shoveled promptly in winter thus normal driving was almost unaffected; the N-FOT trips avoided extreme weather in winter so that the data was biased. Besides, We will also facilitate the model to build a stochastic simulation environment for the testing and evaluation of HAVs.



\section*{Disclaimers}

This work was funded in part by the University of Michigan Mobility Transformation Center Denso Pool project. The findings and conclusions in the report are those of the authors and do not necessarily represent the views of the MTC or Denso.

\addtolength{\textheight}{-12cm}   


\bibliographystyle{IEEEtran}
\bibliography{Mendeley_LeftTurnPaper}

\end{document}